\documentclass[12pt,hyper,letterpaper]{JHEP3}
\usepackage{amsmath, amssymb, amsthm, pictexwd}
\usepackage{dcpic}

\numberwithin{equation}{section}

\def\defeq{\buildrel\rm def\over=}

\def\wdg{{\wedge}}                              

\def\BZ{\mathbb{Z}}

\def\BR{\mathbb{R}}

\def\mcO{\mathcal{O}}
\def\mcA{\mathcal{A}}
\def\mcB{\mathcal{B}}
\def\mcC{\mathcal{C}}
\def\mcD{\mathcal{D}}

\def\id{\mathrm{id}}

\def\tr{\mathrm{tr}}

\def\thh{{\overline{\theta}}}

\def\Ahat{{\widehat{A}}}

\def\CS{{\mathrm{CS}}}

\def\nn{\nonumber}
\def\ie{\textit{i.e.,\ }}

\preprint{MIFP-09-34\\ NSF-KITP-09-142\\}

\title{Geometric Aspects of D-branes and T-duality}

\author{Katrin Becker$^\mathrm{a}$ \\  \email{kbecker@physics.tamu.edu}\\ ${}^{a}$Department of Physics\\Texas A\&M
University\\College Station, TX 77843, USA }

\author{Aaron Bergman$^\mathrm{a,\,b}$  \\   \email{abergman@physics.utexas.edu}\\
${}^{b}$Kavli Institute for Theoretical Physics\\
University of California at Santa Barbara\\Santa
Barbara, CA 93106, USA
 }

\abstract{We explore the differential geometry of T-duality and D-branes.
Because D-branes and RR-fields are properly described via K-theory,
we discuss the (differential) K-theoretic generalization of
T-duality and its application to the coupling of D-branes to
RR-fields. This leads to a puzzle involving the transformation of
the A-roof genera in the coupling.}


\begin{document}

\section{Introduction}

A background for perturbative string theory is given by a  two dimensional conformal field theory. When that field theory possesses a compact Abelian symmetry, it can be gauged. Through a process known as Abelian duality (see, for example, \cite{Giveon:1994fu}), the gauge field can be integrated out resulting in a new background that also possesses a compact Abelian symmetry. This duality is called T-duality. In its simplest form, it exchanges a circle of radius $R$ with a circle of radius $\alpha'/R$. More generally, Buscher has derived transformation rules for nonlinear sigma models in \cite{Buscher:1987qj,Buscher:1987sk,Buscher:1985kb}. One can also see T-duality from the target space point of view by compactifying supergravity on a circle. In that context, the transformation rules have been derived in \cite{Meessen:1998qm,Bergshoeff:1995as}. 

In the above references,\footnote{For another derivation, see \cite{Hassan:1999bv, Hassan:1999mm}.} the transformation of the Ramond-Ramond fields is described in components. It was soon recognized in, for example, \cite{Hori:1999me, Bouwknegt:2003vb}, that these rules can be written in a more geometric form. In particular, T-duality is a duality between two circle bundles over a given base. We can form the fiber product which is a torus bundle over the base whose fiber is the product of the fibers of the two circle bundles. Let us denote the T-dual pair of circle bundles by $M$ and $M'$ with base $\mcB$. Then, the fiber product is written as $M \times_\mcB M'$. The Buscher rules are equivalent to taking the RR-potential on $M$ (chosen to be independent of the fiber direction), multiplying it by a particular form on $M \times_\mcB M'$ and then integrating over the fiber of $M$. This gives a RR-potential on $M'$ which is the T-dual field. 

The simplest version of the coupling of a  D-brane in type IIA or IIB string theory to background RR-fields can be written as follows
\begin{equation}
\label{dcboring}
\int_X e^{B} C\ .
\end{equation}
Here $C$ is the total RR-potential, and $B$ is the NSNS-potential which is globally defined on the brane worldvolume because the Freed-Witten anomaly cancellation condition ensures that the NSNS-field strength $H$ is trivializable on the D-brane. This formula is incomplete in many ways. For example, we could add a gauge field on the D-brane. For a single D-brane, this is accomplished by replacing $B$ by the gauge-invariant combination $B+F$. For a stack of D-branes, however, we have a non-Abelian gauge field, and we need to add a more complicated set of terms due to Myers \cite{Myers:1999ps}. We will ignore these terms here, although their incorporation into this formalism is an interesting puzzle. Instead, we will concentrate on certain $\alpha'$ corrections to \eqref{dcboring}. In particular, these have a topological character and are usually given as\footnote{There are some subtle contributions to this formula relating to spin$_c$ structures and self-duality that we are neglecting.}
\begin{equation}
\label{dcright}
\int_X \sqrt{\frac{\Ahat(TX)}{\Ahat(NX)}} e^B C\ ,
\end{equation}
where $NX$ is the normal bundle to the D-brane, $X$.

The topological nature of these corrections arises because they can be derived by anomaly inflow arguments \cite{Green:1996dd, Cheung:1997az}. It is important to note that, while the total anomaly is given by the integral over a product of characteristic classes and is independent of the choice of forms that represent those classes, as $e^B C$ is not closed, \eqref{dcright} depends on the explicit form of the A-roof genus and \textit{not} just on its cohomology class. Thus, it is an interesting question to ask the specific form of the above corrections. For example, one could form the A-roof genus out of the Levi-Civita connection, an $H$-connection or some other connection on the bundles associated to the tangent and normal bundles to $X$. 

To better understand the coupling \eqref{dcright}, recall that the correct quantization of D-branes \cite{Witten:1998cd} and RR-fluxes \cite{Moore:1999gb} is in terms of K-theory rather than ordinary cohomology. In this language, the brane coupling is given by the pairing of the K-theory class representing the brane with the K-theory class representing the RR-field. This pairing is related to index theory and naturally incorporates the terms involving the A-roof genus. Thus, as opposed to integrating a differential form over the fiber to obtain our T-dual fields, we should instead do this integral in K-theory. A version of this appears in \cite{Hori:1999me} and is exploited extensively in \cite{Bouwknegt:2003vb}. However, life is not quite this simple. It is the RR-charge that is a class in K-theory. The RR-potential is instead part of a class in \textit{differential} K-theory, a generalization of K-theory which not only includes the quantized flux, but also the RR-field itself.\footnote{For a nice introduction to the differential cohomology and differential K-theory used in this paper, please see \cite{Freed:2000ta}.} In addition, because we have a nontrivial $H$-flux, we must use twisted differential K-theory. The goal of this paper is to explore how T-duality of D-branes, RR-fields and their couplings can all be viewed in light of the geometrical T-duality transformation. We will find various conditions that lead to invariance of the D-brane coupling under T-duality and discuss how they relate to the K-theoretic description of D-branes and RR-fields. As we discuss in section \ref{sec:verif}, these relations are physically mysterious. While they are useful mathematically, they are surely not the complete story for the physics and may not even be correct in that context. It is interesting to note, however, that from two somewhat different directions, we arrive at the condition that, when a geometric background has a free Abelian symmetry, the ratio of A-roof genera in \eqref{dcright} is a form pulled back from the base of the circle bundle. Something akin to this has been shown to hold in some examples with $\mathcal{N}=2$ supersymmetry in the heterotic string \cite{Becker:2009zx}. This obviously has consequences towards the ambiguity discussed in the previous paragraph. We leave the proper interpretation of these conditions as a puzzle.

This paper is organized as follows. The first three sections are devoted to differential geometric aspects of T-duality. Much of this material is well-known, but we hope that a unified presentation in a geometric language will be helpful for the reader. In section \ref{sec:setup}, we rewrite the Buscher rules in a geometric form and give the geometry of T-dual pairs of D-branes. In section \ref{sec:gauge}, we see how the $B \leftrightarrow F$ gauge invariance behaves under T-duality. In section \ref{sec:selfdual}, we show that the Buscher rules preserve the self-duality constraint on the RR-fields paying special attention to the signs that arise. The remainder of the paper is devoted to the K-theoretic aspects of T-duality. In section \ref{sec:ktheory}, we recall how the coupling \eqref{dcright} can be interpreted as a pairing in K-theory, and we give a K-theoretic version of the T-duality transformation. In section \ref{sec:verif}, we derive certain conditions under which T-duality preserves the brane coupling, and we discuss the import of these conditions. In section \ref{sec:index}, we show that, if the Buscher rules are uncorrected, the pairing in K-theory is preserved by T-duality. Finally, in section \ref{sec:branes}, we derive a condition such that the geometric T-duality transformation can be applied to branes to give the expected results. Combined with the results of sections \ref{sec:ktheory} and \ref{sec:branes}, this gives another demonstration of the invariance of the D-brane coupling.

\section{The setup}\label{sec:setup}

We will begin by making precise the relation between the Buscher rules and the integration over the fiber product described in the introduction. Our spacetime $M$ is a circle bundle over $\mcB$ with metric
$$
ds_M^2 = ds^2_\mcB + e^{2\phi(b)} (d\theta + A)^2\ .
$$
Thus, $\mcA = d\theta + A$ is a connection on the circle bundle $M$,
and $d\mcA = e$ where $e = e(M)$ is the Euler class of $M$. We also
have an $H$-flux given by
$$
H = - \gamma+ \mcA \wdg \lambda\ ,
$$
where $\lambda$ is a closed two-form on $\mcB$, and $\gamma$ is a three-form on $\mcB$ such that $d\gamma = e \wdg \lambda$, implying that $H$ is closed.

The T-dual geometry is a new circle bundle $M'$ over $\mcB$ with
Euler class $\lambda$. As a general rule, we will use primes to
denote T-dual quantities. To use the Buscher rules, we first note
that locally
$$
A_\mu=\frac{G_{\mu\theta}}{G_{\theta\theta}}\ .
$$
Similarly, we can
locally trivialize
$$
H = dB\qquad  {\rm with} \qquad B = \beta - \mcA \wdg \alpha\ .
$$
In components, we have
$$
B_\parallel= \beta - A \wdg \alpha\ , \qquad B_{\theta} = \alpha\ ,
$$
where we have set $B=B_\parallel + B_\theta \wedge d \theta$ with
$B_\parallel$ the component of $B$ along the base, $\mcB$, and $B_\theta$
the component along the fiber. This implies that
$$
\lambda = d \alpha\ , \qquad \gamma= e \wedge \alpha
-d \beta\ .
$$

Writing the coordinate for the T-dual circle bundle as $\thh$, the Buscher rules give the T-dual metric as
\begin{equation}
ds^2_{M'} = ds^2_\mcB + e^{-2\phi(b)}(d\thh + A')^2\ ,
\end{equation}
where the (locally defined) one-form $A' = B_\theta$. Thus, we have the (globally defined) T-dual connection $\mcA' = d\thh + A' = d\thh + \alpha$ which satisfies $e(M') = d\mcA' = \lambda$ as expected. We also have that
\begin{equation}
\label{btrans}
B'_\parallel =\beta\ , \qquad B'_{\bar \theta} = A\ ,
\end{equation}
since $B'_\parallel = B_\parallel - B_\theta \wdg A$.
This tells us that the local version of the T-dual B-field is
\begin{equation}
\label{localbt}
B' = \beta + A \wdg d\thh = B_\parallel + A \wdg \mcA'\ ,
\end{equation}
and
$$
H' = dB' = d\beta + dA \wdg d\thh  = d\beta - e \wdg \alpha + e \wdg \mcA' = e \wdg \mcA' - \gamma\ .
$$
It is important here that the T-duality transformation rules act on
the B-field and not just on its flux. Even though we have worked
locally, trivializing both $H$ and the circle bundle $M$, it is
possible to define this transformation in an invariant manner using
differential cohomology \cite{dfpcomm}. To summarize, the T-dual
NSNS-fields are
\begin{equation}
\label{rulessumm}
\begin{split}
& A'  = B_\theta\ ,\\
& \phi'=-\phi\ , \\
 & B'= B_\parallel +
A \wdg \mcA'\ .
\end{split}
\end{equation}

In order to define a T-duality transformation on the RR-fields, we form the fiber product $M \times_\mcB M'$ which is defined to be the set of all $(m,m') \in M \times M'$ such that $m$ and $m'$ project to the same point in $\mcB$.  This is a torus bundle over the base, $\mcB$. Let $C$ be the differential form representing the sum of all the RR-potentials. By assumption, it has no $\theta$ dependence. The T-duality formulae for RR-potentials given in \cite{Myers:1999ps} are equivalent to
\begin{equation}
\label{tdwrong}
C' = \pi_{2*}\left(\pi_1^*C \wdg e^{\mcA' \wdg \mcA}\right)\ ,
\end{equation}
where $\pi_1$ and $\pi_2$ are the projections from $M \times_B M'$ to $M$ and $M'$ respectively.
This expression appears in \cite{Bouwknegt:2003vb}, building on the work in \cite{Hori:1999me}.
We can derive the usual form of the Buscher rules as follows. First, we locally write the RR-potential $C$ as
$$
C = C_\parallel + C_\theta \wdg d\theta\ .
$$
In the expression \eqref{tdwrong}, the pushforward $\pi_{2*}$
represents an integral over the variable, $\theta$, and thus we can
write things as follows
$$
C' = \int _\theta \left[ C(1 +  (d\thh + \alpha)(d\theta +
A))\right] = C_\theta + C_\parallel (d\thh + \alpha) - C_\theta A
(d\thh + \alpha)\ ,
$$
or
\begin{equation}
\label{wtdualcomp} C'_\thh = C_\parallel - C_\theta \wdg A\ ,
\qquad C'_\parallel = C_\theta + C_\parallel \wdg A'+ C_\theta \wdg
A' \wdg A\ .
\end{equation}
These are precisely the same rules as given in \cite{Myers:1999ps}. It is straightforward to verify that this transformation squares to the identity.

Next, we wish to add a D-brane to this setup. We will choose it so that it wraps the fiber of $M$. Let $X$ be a submanifold of $\mcB$, and $i : X \to \mcB$ be the embedding. We define the D-brane $X_M = i^*(M)$ to be the restriction of the bundle $M$ to $X$, and we have the embedding $i_M : X_M \to M$. Thus, we have the following diagram of spaces:
\begin{equation*}
\begin{split}
\begindc{\commdiag}[4]
\obj(15,10)[X]{$X$}
\obj(30,10)[B]{$\mcB$}
\obj(5,20)[XM]{$X_M$}
\obj(20,20)[M]{$M$}
\obj(40,20)[Mp]{$M'$}
\obj(30,30)[M2]{$M \times_\mcB M'$}
\mor{XM}{X}{}
\mor{XM}{M}{$i_M$}
\mor{M}{B}{$\rho_1$}
\mor{X}{B}{$i$}
\mor{M2}{M}{$\pi_1$}
\mor{M2}{Mp}{$\pi_2$}
\mor{Mp}{B}{$\rho_2$}
\enddc
\end{split}\ .
\end{equation*}

The picture to keep in mind here is that the D-brane $X_M$ is a submanifold of $M$ wrapping both $X$ and the fiber of $M$ over $X$. It must obey the Freed-Witten anomaly equation, $i_M^*(H) = dB$. Let $\alpha = \pi_{1*} B$. Then $d\alpha = \pi_{1*} dB = i^*\lambda$. Now let $\beta = B + i_M^*\mcA \wdg \alpha$. This only has components along the base, and $d\beta = i^*e \wdg \alpha  - i^*\gamma$. These are exactly the pullbacks of the local form for $B$ given above, but since we restricting to the worldvolume of the brane, the Freed-Witten condition means that $\alpha$ and $\beta$ are now globally defined.

The T-dual brane does not wrap the fiber of $M'$, and thus it represents a section of the bundle $M' \to \mcB$ restricted to the submanifold $X$. More precisely, this means that there exists a section of the bundle $i^*(M')$ which we write as a map $s : X \to  i^*(M')$. This trivializes $i^*(M')$ which means that $i^*(e(M')) = i^*\lambda$ must be exact. This is indeed true as from above we have that $i^*\lambda = d\alpha$. Next, let $\eta$ be the embedding of $i^*(M') \to M'$. Then there is a map $i_{M'} = \eta \circ s : X \to M'$ which is the embedding of the T-dual D-brane. We choose the section, $s$, such that the wrapping number over any nontrivial loop in the base is zero. This is dual to there being no gauge flux in the fiber direction on the brane $X_M$.\footnote{One way to see this is that a wrapping of the brane around the coordinate $\thh$ can be replaced by the addition of a cohomologically non-trivial one-form to the metric, \ie $(d\thh + A')^2 \rightarrow (d\thh +A' +  \kappa)^2$. Under T-duality, this comes from adding $\kappa$ to $B_{\mu\theta}$ which can be exchanged for a gauge field under the $B\leftrightarrow F$ gauge transformation. This gauge invariance means that we should really never talk about $B$ alone. See the next section for more details.} Thus, we can choose coordinates such that the section, $s$, is at constant $\thh$ and, hence, $i_{M'}^*(\mcA') = \alpha$. To see the Freed-Witten condition, note that $i^*_{M'}(H') = i^*_{M'}(e) \wdg \alpha - i^*_{M'}\gamma = i^*e \wdg \alpha - i^*\gamma = d\beta$, which means that the B-field on the brane is given by $B' = \beta$. This is precisely the pullback of the local expression \eqref{localbt} to the brane worldvolume as expected.

Let us summarize the results of this section. Our original spacetime is denoted by $M$ and our T-dual spacetime is given by $M'$. Each is a circle bundle over a base, $\mcB$, with connections $\mcA$ and $\mcA'$, respectively. The Euler classes are given by
$$
e(M) = d\mcA = e \quad \mbox{and} \quad e(M') = d\mcA' =  \lambda\ ,
$$
and the H-fluxes are given by
$$
H = \mcA \wdg \lambda - \gamma \quad \mbox{and} \quad H' = e \wdg \mcA' - \gamma\ .
$$
We have a brane $X_M$ embedded into $M$ which wraps the fiber of $M$ and a T-dual brane given by $X$ embedded into $M'$. Their components along the base, $\mcB$, are given by the same embedding of $X$ into $\mcB$. The B-fields on the worldvolumes of the branes are given by
$$
B = \beta - \mcA \wdg \alpha \quad \mbox{and} \quad  B' = \beta\ .
$$
These satisfy $dB = H$ and $dB' = H'$ where we restrict $H$ and $H'$ to the worldvolumes of the branes. This gives relations between $\alpha$, $\beta$, $e$ and $\lambda$.

\section{Gauge invariance}\label{sec:gauge}

In the previous section, we set $F=0$ on the brane. This is not a gauge invariant statement, however, as there exists a symmetry
\begin{equation}
\label{bfgauge}
F \to F + \Lambda\ , \qquad B \to B - \Lambda\ ,
\end{equation}
with $d\Lambda = 0$. In particular, in the brane coupling, we replace $B$ by the gauge invariant quantity $B+F$. For example, \eqref{dcright} is replaced by
$$
\int_X \sqrt{\frac{\Ahat(TX)}{\Ahat(NX)}} e^{B+F} C\ ,
$$
and similarly for the other forms of the coupling discussed later. In this section, we will see how this symmetry is implemented in the context of T-duality. We will only consider Abelian gauge fields as non-Abelian degrees of freedom necessitate the inclusion of Myers terms \cite{Myers:1999ps}.

To begin with, let us examine the effects of the addition of a gauge flux, $F$. We can decompose $F = F_\parallel + F_\theta d\theta$. The rules for T-duality give that the T-dual field strength is
\begin{equation}
\label{tdualf}
F' = F_\parallel\ .
\end{equation}
The effect of $F_\theta$, then, is that it changes the position of the T-dual D-brane. In particular, recall that the position of the dual brane is given by the value of the Wilson loop around the circle
$$
\exp\left(2\pi i \int d\theta A_\theta\right)\ .
$$
Thus, if $F_\theta = df$ for some $\BR$ valued function, $f(b)$ on $\mcB$, then $A_\theta = f$, and we have that the dual position of the brane is given by $f\!\! \mod 1$ (where we have set the circumference of the dual circle to be 1).

To address the situation where $F_\theta$ is not exact, recall that the relation $F_\theta = df$ is equivalent to $F_\theta = f^*(d\phi)$ where $\phi$ is a coordinate on $S^1$. In fact, we can write any closed one form as the pullback of $d\phi$ for some function $f : \mcB \to S^1\cong \BR/\BZ$. This function provides the position of the T-dual brane. More precisely, recall from the previous section that we have an embedding $i_{M'} : X \to M'$. This map trivializes the circle bundle $M'$ when restricted to $X$. Thus, we can think of $i_{M'}$ as a map from $X$ into $X \times S^1$ given by $i_{M}(x) = (x,0)$ where, as above, we parametrize $S^1$ by $\BR/\BZ$. This is the location of the T-dual D-brane when there is no $F_\theta$ flux. The addition of a flux $F_\theta = f^*(d\phi)$ lets us define a new function
$$
i^f_M(x) = (x,f)\ ,
$$
which is the embedding of the T-dual D-brane. This D-brane has a flux on it given by $F' = F_\parallel$.

We can now understand the gauge invariance \eqref{bfgauge}. We make the decomposition $\Lambda = \Lambda_\parallel + \Lambda_\theta d\theta$. We begin with the case $\Lambda_\theta = 0$. This leaves $B_\theta$ and $F_\theta$ unchanged, while $B_\parallel$  and $F_\parallel$ are shifted by $\Lambda_\parallel$. From \eqref{btrans} and \eqref{tdualf}, we have that
$$
(B' + F')_\parallel = (B+F)_\parallel\ .
$$ 
Thus, the coupling is invariant under this transformation.

Now let us consider the case where $\Lambda = \Lambda_\theta d\theta$. As above, since $d\Lambda_\theta=0$, we can define a function $f_{\Lambda_\theta} : \mcB \to S^1$ satisfying 
$$
f_{\Lambda_\theta}^*(d\phi) = \Lambda_\theta\ .
$$
Since we have shifted $F\to F + \Lambda_\theta d\theta$, the above discussions tells us that the embedding of the T-dual brane is shifted to $i_{M'}^{f+f_{\Lambda_\theta}}(x) = (x,f + f_{\Lambda_\theta})$. As we also have shifted $B\to B-\Lambda d\theta$, from \eqref{rulessumm}, as $A' = B_\theta$ we have that $A'$ is shifted as
$$
A'\to A' - \Lambda_\theta\ .
$$
This means that the new T-dual metric is
$$
ds^2 = ds_\mcB^2 + e^{-2\phi}(d\thh + A' - \Lambda_\theta)^2\ .
$$
Since the original transformation of $F$ and $B$ was a gauge invariance, we should be able to recover the original T-dual brane and metric via a gauge invariance on the T-dual side. To do so, we make the following coordinate transformation
$$
\sigma = \thh - f_{\Lambda_\theta}\ .
$$
This changes the metric to
$$
ds^2 = ds_\mcB^2 + e^{-2\phi}(d\sigma + A')^2\ .
$$
In these new coordinates, the brane embedding is given by $i_{M'}^{f+f_\Lambda}(x) = (x,f + f_{\Lambda_\theta}-f_{\Lambda_\theta}) = (x,f)$. But these are precisely the original T-dual metric and brane embedding. Thus, we see that the gauge invariance \eqref{bfgauge} is mapped under T-duality to a coordinate transformation, and everything is invariant.

The careful reader will have noticed, however, that the transformation rule for $F$ here is not well-defined because the form $F_\parallel + F_\theta d\theta$ depends on a local trivialization of $X_M$ over $X$. To make this dependence precise, let $s$ be a local section of $X_M$ over some open set, $U \subset X$. Then we define $F^s_\parallel = s^*F$ and $F^s_\theta = \rho_{1*}(F)$. In a new trivialization $t$, we can similarly define $F^t_\parallel$ and $F^t_\theta$. The difference $s-t$ defines a map from $U$ to $U(1)$, and, as above, a one-form $\nu = (s-t)^*(d\phi)$. Then we have $F^t_\parallel - F^s_\parallel = F^s_\theta \wdg \nu$ and $F^t_\theta = F^s_\theta$. Since $F' = F_\parallel$, we have that $\Delta F' =(F^s)' - (F^t)' = F_\theta \wdg \nu$. Thus, it appears that the change in trivialization has led us to a different T-dual gauge flux. However, recall the definition of the T-dual B-field \eqref{localbt}
$$
B' = \beta + A \wdg d\thh\ .
$$
Not only is this expression only defined locally on $X_M$, the fact that it involves $A$ means that it also depends on an explicit trivialization $s : X \to X_M$ such that $A^s = s^*(\mcA)$. In the trivialization $t$, we have $A^t = A^s + \nu$. Thus, the T-dual B-field shifts by $\Delta B' = \nu \wdg d\thh$. It follows from the above discussion that $i^*_{M'}(d\thh) = F_\theta$, and $i^*_{M'}(\Delta B') = \nu \wdg  F_\theta$. Thus, we see that the change in trivialization is precisely a gauge transformation of the form \eqref{bfgauge} on the T-dual side. In fact, this is very much the same as the above situation where the choice of a trivialization masqueraded as a coordinate transformation.

The moral of this story is that the traditional concept of a `gauge field' on the D-brane is not an invariant notion even in when the $H$-flux is trivial.\footnote{Formally, the B-field is an element in the differential cohomology group $\check{H}^3(M)$. The Freed-Witten anomaly cancellation condition states that the pullback of this element to the brane is trivial. The correct notion of an Abelian `gauge field' on the brane is as an explicit trivialization of this class.} One should only speak of gauge invariant quantities, and $F$ is not gauge invariant. Of course, neither is $B$, but to save space, for the rest of the paper we will write $B$ even though we mean the gauge invariant quantity $B+F$.

\section{Self-duality}\label{sec:selfdual}

Self-duality imposes a significant constraint on the RR-field strengths. In this section, we will show that the Buscher rules preserve this constraint.\footnote{One might expect that this restricts the form of the possible corrections to the Buscher rules. However, one should be very careful when discussing the quantization of a self-dual field. For a discussion of the relevant subtleties, see \cite{Belov:2006xj}.} Away from the sources, one can define this field strength as $F = dC + H \wdg C$. It is straightforward to verify that the Buscher rules give
\begin{equation}
\label{fluxtrans}
F' = \pi_{2*}\!\left(\pi^*_1(F)  e^{\mcA' \wdg \mcA}\right)\ .
\end{equation}

Now, let us consider the self-duality constraint in the form presented in \cite{Belov:2006xj}. In Lorentzian signature, one first defines the total RR-flux
\begin{eqnarray*}
IIA : F &=& F_0 + F_2 + F_4 + \star F_4 - \star F_2 + \star F_0\ , \\
IIB : F &=& F_1 + F_3 + F_5 + \star F_3 - \star F_1\ ,
\end{eqnarray*}
with ${\star F_5} = - F_5$ in IIB. These fluxes obey the self-duality relation
$$
F = (-)^\frac{10+\kappa - d}{2} {\star F}\ ,
$$
where $d$ is the degree of the form and $\kappa= d \!\mod 2$. Thus, $\kappa = 0,1$ for IIA and IIB respectively. In what follows, whenever $d$ occurs in an expression such as the above, it is the degree of the form following the sign, extended by linearity to act on arbitrary forms. For example, given a $p$ form $\omega_p$ and a $q$ form $\omega_q$, $(-)^d(\omega_p + \omega_q) = (-1)^p\omega_p + (-1)^q \omega_q$.

We want to write a formula for the Hodge star on an arbitrary circle bundle $M$ with the metric
$$
ds_M^2 = ds^2_\mcB + e^{2\phi}(d\theta + A)^2\ .
$$
We choose an orthonormal frame of one-forms $e^a$ on $\mcB$ and add $e^\vartheta = e^\phi (d\theta + A) = e^\phi \mcA$ to complete the frame on $M$. Note that, despite the regrettable collision of notation, $e^\phi$ here is a function on $\mcB$ and not a one-form. Recall that the Hodge star is defined to satisfy
\begin{equation}
\label{hodgedef}
\int_M \sqrt{|\mathrm{det}(g)|} \alpha_{\mu_1 \dots \mu_n} \beta^{\mu_1 \dots \mu_n} = \int_M \alpha \wdg {\star \beta}\ .
\end{equation}
If we write
$$
\alpha = \alpha_\wr + \alpha_\theta \wdg e^\phi \mcA\ ,
$$
we have that, in orthonormal components,
$$
\alpha_{a_1 \dots a_n} = \alpha_\wr \quad \mathrm{and} \quad \alpha_{a_1 \dots a_{n-1} \vartheta} = \alpha_\theta\ .
$$
Thus, choosing a mostly plus signature,
\begin{equation}
\label{hodgedual1}
\begin{split}
\int_M \sqrt{|\mathrm{det}(g)|} \alpha_{\mu_1 \dots \mu_n} \beta^{\mu_1 \dots \mu_n} &=
2\pi \int_B e^\phi\left( (\alpha_\wr\cdot \beta_\wr)+ (\alpha_\theta \cdot \beta_\theta) \right) \\
&= 2\pi \int_B e^\phi \left( \alpha_\wr \wdg {\star_\mcB \beta_\wr} + \alpha_\theta \wdg {\star_\mcB \beta_\theta} \right)\ ,
\end{split}
\end{equation}
where
$$
(\alpha_\wr \cdot \beta_\wr) = \sqrt{|\mathrm{det}(g_\mcB)|}\alpha_{\wr\mu_1 \dots \mu_n} \beta_\wr^{\mu_1\dots\mu_n}\ ,
$$
and similarly for $(\alpha_\theta \cdot \beta_\theta)$.

The right side of \eqref{hodgedef} can be written
\begin{equation}
\label{hodgedual2}
\begin{split}
\int_ M\alpha \wdg {\star\beta} &= \int_M (\alpha_\wr + \alpha_\theta \wdg e^\phi \mcA) \wdg ((\star\beta)_\wr + (\star\beta)_\theta \wdg e^\phi \mcA) \\
&= \int_M e^\phi \left(\alpha_\wr \wdg (\star\beta)_\theta +  \alpha_\theta \wdg (-)^d(\star\beta)_\wr\right) \wdg \mcA \\
&= 2\pi \int_B e^\phi \left(\alpha_\wr \wdg (\star\beta)_\theta + \alpha_\theta \wdg  (-)^{d}(\star\beta)_\wr\right)\ .
\end{split}
\end{equation}
Comparing \eqref{hodgedual1} and \eqref{hodgedual2}, we obtain
$$
\star\beta = (\star\beta)_\wr + (\star\beta)_\theta \wdg e^\phi \mcA = (-)^d {\star_\mcB\beta_\theta} + {\star_\mcB\beta_\wr} \wdg e^{\phi} \mcA\ .
$$
Returning to the case where $M$ is Lorentzian and ten dimensional, the self-duality constraint on $F = F_\wr + F_\theta \wdg e^\phi \mcA$ is that
\begin{equation}
\label{selfdual1}
F_\wr = (-)^\frac{2 + \kappa + d}{2} \star_\mcB\!F_\theta\ ,
\end{equation}
and, equivalently,
\begin{equation}
\label{selfdual2}
F_\theta = (-)^\frac{1 + \kappa - d}{2} \star_\mcB\!F_\wr\ .
\end{equation}

The T-dual of $F$ is
\begin{equation}
\label{tdcomp}
\begin{split}
F' &= \pi_{2*}\left((F_\wr + F_\theta \wdg e^\phi \mcA)\wdg(1 + \mcA' \wdg \mcA)\right) \\
&= F_\wr \wdg \mcA' + e^\phi F_\theta \\
&= e^\phi (F_\theta + F_\wr \wdg e^{-\phi} \mcA')\ .
\end{split}
\end{equation}
Hence
\begin{equation}
\label{tdualcomp}
F'_\wr = e^\phi F_\theta \quad\mathrm{and}\quad F'_\theta = e^\phi F_\wr\ .
\end{equation}
To verify self-duality, we compute
$$
F'_\wr = e^\phi F_\theta =  (-)^\frac{1 + \kappa - d}{2} e^\phi \star_\mcB\!F_\wr =  (-)^\frac{1 + \kappa - d}{2} {\star_\mcB F_\theta'} = (-)^{-(\kappa' + d)} (-)^\frac{2 + \kappa' + d}{2} {\star_\mcB F_\theta'}\ ,
$$
where we have used that $\kappa' = 1 - \kappa$. Since the degrees of $\star_\mcB F'_\theta$ are the same as the degrees of $F'$, $\kappa' = d \!\mod 2$ and the first sign is 1. Thus, we recover the self-duality constraint \eqref{selfdual1} for the T-dual field strength.

\section{K-theory and RR-fields}\label{sec:ktheory}

In this section, we begin our discussion of the K-theoretic aspects of T-duality. First, we will `derive' the expression \eqref{dcright} for the brane coupling as follows.\footnote{A variant of this argument was, in fact, the original motivation for the relation of D-brane charges to K-theory and appears in \cite{Minasian:1997mm}. The derivation here is well-known.} We will briefly depart from the notation of section \ref{sec:setup} and discuss a brane $X$ embedded into a manifold $M$ by the map $i_M$. Since we know that D-branes and RR-fields naturally live in twisted differential K-theory, let us write $i_{M*}(1)$ for the push-forward of the fundamental class of $X$ to $M$. This is possible in twisted K-theory because of the Freed-Witten condition. The twist is provided by an element in differential $H \in \check{H}^3(M)$ \cite{dfpcomm}, and we write the twisted differential K-theory group as $\check{K}^\bullet_H(M)$ . Physically, this is just the fact that we have to keep track of the B-field and not just its flux. We will denote the class of the RR-potential by $\mcC$. From Moore and Witten \cite{Moore:1999gb}, we have that the differential form version of the RR-potential is
\begin{equation}
\label{fmchern}
C = \sqrt{\Ahat(M)} \mathrm{ch}(\mcC)\ ,
\end{equation}
and
$$
F = d_H C \defeq dC + H \wdg C\ .
$$
The coupling is given by the K-theoretic integral
\begin{equation}
\label{ktint}
\int_M i_{M*}(1) \cdot \mcC\ .
\end{equation}
The integrand lives in the differential K-group $\check{K}^1(M)$, and the integral takes this to $\check{K}^1(pt) \cong \BR / \BZ$ which we can exponentiate and add to the action. The integral in differential K-theory is related to the index of the Dirac operator and requires a Riemannian structure on $T\!M$.

To write this in cohomology, we use the index theorem of \cite{Freed:2009it} to obtain\footnote{Like all the formulae in this paper involving the RR-potential, this formula is meant to be impressionistic. It can be made sense of by unravelling the definition of the pushforward in differential K-theory.  However, we are neglecting, among other things, the self-duality constraint and the issue of Spin$_c$ structures. See \cite{Freed:2000ta} and \cite{Freed:2009it} for more details.}
\begin{equation}
\label{indthm}
\int_M \Ahat(M) \mathrm{ch}(i_{M*}(1)) \wdg \mathrm{ch}(\mcC)\ .
\end{equation}
The Riemann-Roch theorem\footnote{To my knowledge, this remains unproven for twisted differential K-theory. For ordinary twisted K-theory, the theorem is proven in \cite{Carey:2007dt}. As always, we are neglecting many issues here and making little pretense towards rigor, leaving that for other work.} roughly gives that
\begin{equation}
\label{rroch}
\mathrm{ch}(i_{M*}(1)) = i_{M*}\left(\Ahat(N(X/M))^{-1}e^{B}\right)\ .
\end{equation}
In the untwisted case, the Riemannian structure on the normal bundle gives rise to certain connections that determine the A-roof genus. Since we do have a twist, however, we will leave the A-roof genus unspecified. The right side of \eqref{rroch} should be interpreted as a current supported on $X$. Inserting this into the above expression and using the projection formula, we obtain
\begin{equation}
\label{coup1}
\int_{X} \frac{1}{\Ahat(N(X/M))} e^{B}\  i^*_M\!\!\left(C \sqrt{\Ahat(M)}\right)\ .
\end{equation}
For a D7 brane, we can write this in terms of Pontryagin classes as follows (we set $B=0$ for convenience):
\begin{equation}
\begin{split}
\int_{D7} i^*_M(C_8) &+ \frac{1}{24}\left(p_1(N(X/M)) - \frac{1}{2}i^*_M(p_1(TM))\right) \wdg i^*_M(C_4) \\
&+ \frac{1}{1440}\left(p_2(N(X/M)) - \frac{1}{2}i^*_M(p_2(TM))\right)\wdg i^*_M(C_0) \\
&+ \frac{1}{640}\left(\frac{1}{3}p_1(N(X/M))^2 + \frac{1}{4}i^*_M(p_1(TM)^2)\right)\wdg i^*_M(C_0)\ .
\end{split}
\end{equation}
To write a Pontryagin class in terms of forms, we have to choose a connection on the relevant bundle. If we denote its curvature as $F$, we have
$$
p_1(V) = -\frac{1}{8\pi^2} \tr(F^2)\ , \qquad p_2(V) = \frac{1}{128 \pi^4}\left(\tr(F^2)^2 - 2\tr(F^4)\right)\ .
$$

Because the pullback of $TM$ to $X$ decomposes as the direct sum $TX \oplus N(X/M)$, on the level of cohomology, the pullback of $[\Ahat(M)]$ factors as $[\Ahat(TX)] \cup [\Ahat(N(X/M))]$. If we take this to hold on the level of forms, we can turn \eqref{coup1} into the expression \eqref{dcright}
\begin{equation}
\label{coup2}
\int_{X} \sqrt{\frac{\Ahat(TX)}{\Ahat(NX)}} e^{B} C\ .
\end{equation}
However, since $e^B C$ is not closed, even though the two ratios of genera are cohomologous, it does not mean that \eqref{coup1} and \eqref{coup2} are equal. Since the difference between the ratio of genera is an exact form, we can write it as $d\CS$, and we have that
\begin{equation}
\label{coupdiff}
\eqref{coup1} - \eqref{coup2} = \int_{X} d\CS\ e^B C = \int_X \CS\ e^B d_H C = \int_X \CS\ e^B F\ . 
\end{equation}
We can, in principle, write down a version of the brane coupling for any representative in the cohomology class of the ratio of $\Ahat$-genera. The difference will always amount to a choice of $\CS$ in \eqref{coupdiff}. We should emphasize that there are two separate but related issues here. The first is the form of the coupling in terms of the A-roof genera. This is the difference between, say, \eqref{coup1} and \eqref{coup2}. The second is the fact that we have not specified how to form the relevant A-roof genera. Each of these issues leads to a difference in the form of \eqref{coupdiff}. For the first issue, K-theory suggests that the form \eqref{coup1} is most natural, but it tells us less about the second issue. These choices may be related to the `extra' terms in the brane coupling found by Craps and Roose \cite{Craps:1998tw}.

Because these objects live in K-theory we must extend in some way the Buscher rules \eqref{tdwrong} . Returning to the setup of section \ref{sec:setup}, we can suggest the K-theoretic formula
\begin{equation}
\label{tdkright}
\mcC' = \pi_{2*}(\Theta(\pi_1^*\mcC))
\end{equation}
as an obvious refinement of \eqref{tdwrong}. This formula appears for non-differential K-theory in \cite{Bouwknegt:2003vb}. Here $\Theta$ is an isomorphism between $\check{K}_{H}(M \times_\mcB M')$ and  $\check{K}_{H'}(M \times_\mcB M')$ given by a canonical isomorphism of the pullbacks of the B-fields on $M$ and $M'$ as elements of $\check{H}^3(M \times_\mcB M')$ \cite{dfpcomm}. This is related to the canonical trivialization $H - H' = d(\mcA' \wdg \mcA)$. To obtain an expression in terms of cohomology, we take Chern characters giving
\begin{eqnarray}
\label{comptdual}
C' &= &\sqrt{\Ahat(M')} \mathrm{ch}(\mcC') = \sqrt{\Ahat(M')} \mathrm{ch}(\pi_{2*}(\Theta(\pi_1^*\mcC)))\nn\\
&=&\sqrt{\Ahat(M')} \pi_{2*}\left(\Ahat\left(T\!\left(\frac{M \times_\mcB M'}{M'}\right)\right) \pi_1^* \mathrm{ch}(\mcC) e^{\mcA' \wdg \mcA}\right) \\
&= &\pi_{2*}\left(\sqrt{\frac{\Ahat(M')}{\Ahat(M)}} \Ahat\left(T\!\left(\frac{M \times_\mcB M'}{M'}\right)\right) e^{\mcA' \wdg \mcA} \pi_1^*C\right)\ .\nn
\end{eqnarray}
Here, we have used the index theorem in differential K-theory \cite{Freed:2009it}.\footnote{As with the Riemann-Roch theorem, this theorem is for nontwisted differential K-theory, but we are assuming it holds for the twisted case. The pushforward in ordinary (not differential) twisted K-theory is given in \cite{Carey:2005mx}. Again, this formula is impressionistic and should be interpreted along the lines of \eqref{indthm}. However, it becomes precise when applied to field strengths.} This bundle is canonically isomorphic to that of the projection from $M$ to $B$, so we can write the above expression as
\begin{equation}
\label{tdright}
C' = \pi_{2*}\left(\sqrt{\frac{\Ahat(M')}{\Ahat(M)}} \Ahat\left(T(M/\mcB)\right) e^{\mcA' \wdg \mcA} \pi_1^*C\right)\ .
\end{equation}
Again, since we are concerned with particular representatives and not just cohomology classes, we must still determine the A-roof genera to properly evaluate this expression. For an arbitrary choice of the A-roof genera, this appears to be a correction to the Buscher rules for T-duality. The significance of this will be discussed in the next section.

\section{The invariance of the couplings}\label{sec:verif}

To verify invariance of the brane couplings, we will first write them in the manner of \eqref{coup1}:
\begin{equation}
\label{xmcouple}
\int_{X_M} \frac{\sqrt{i^*_M\Ahat(M)}}{\Ahat(N(X_M/M))} e^{\beta} e^{\alpha \wdg \mcA} i_M^*(C)\ .
\end{equation}
Similarly, we can define $C'$ and write the coupling for the brane $X$ as
\begin{equation}
\label{xcouple}
\int_{X} \frac{\sqrt{i^*_{M'}\Ahat(M')}}{\Ahat(N(X/M'))} e^{\beta} i_{M'}^*(C')\ .
\end{equation}
As noted in the previous section, these expressions differ from the expression \eqref{dcright} given in the introduction.

Because T-duality is its own inverse,\footnote{This is nontrivial to prove in the K-theory case, and we will not attempt to do so here.} to verify that the coupling is invariant, it suffices to check only one direction. Thus, we will manipulate \eqref{xcouple} so that it is equal to \eqref{xmcouple}. To begin with, we plug in the T-duality equation \eqref{tdright}
\begin{equation}
\int_{X} \frac{\sqrt{i^*_{M'}\Ahat(M')}}{\Ahat(N(X/M'))}  e^{\beta}\, i^*_{M'}(C') =  \int_X\frac{i^*_{M'}\Ahat(M')}{\Ahat(N(X/M'))} e^{\beta}\, i^*_{M'}\pi_{2*}\left(\frac{\Ahat\left(T(M/\mcB)\right)}{\sqrt{\Ahat(M)}} e^{\mcA' \wdg \mcA} \pi_1^*C\right)\ .
\end{equation}


Before proceeding, it is worth describing the picture behind the following calculation. Recall that our brane $X_M$ wraps the fiber of $M$ over $X$. The T-dual brane, on the other hand, does not wrap the fiber of the T-dual space $M'$. Now, our expression for the T-dual RR-fields expresses them as an integral over the fiber of $M$; that is the meaning of the push-forward $\pi_{2*}$. On the other hand, the expression for the brane coupling on the T-dual side involves the integral of the T-dual RR-fields over the worldvolume of the brane, $X$. But since the T-dual RR-fields are given by integrals over the fiber of $M$, the integral in the coupling can be expressed as an integral over $X_M$ just done in two parts: first over the fiber to obtain the T-dual RR-field, and then over $X$ to obtain the coupling. However, $X_M$ is precisely the worldvolume of our original brane. Thus, all that remains to do is to compare the explicit expressions over $X_M$. In particular, that is the content of the formula \eqref{tdbc} below: it is a rewriting of the coupling for the T-dual brane, \eqref{xcouple}, and we wish to compare it to the coupling \eqref{xmcouple}. The factor $e^{\mcA' \wdg \mcA}$ corrects the B-field on the brane, and the two brane couplings agree provided me make a number of compatibility assumptions about the A-roof genera.

To make this more precise, recall that $M \times_\mcB M'$ is a circle bundle over $M'$ which we can pull back to $X$ by $i_{M'}$. The resulting space is
$$
X \times_{M'} (M \times_\mcB M') \cong X \times_\mcB M = X_M\ .
$$
This can be seen intuitively by nothing that we are pulling back the fiber of $M \times_\mcB M'$ to the base by the section of $M'$ over $X$. This is precisely the fiber of $M$ over the image of $X$. Explicitly, the space on the left is the collection of triples $(x,m,m')$ such that $i_{M'}(x) = m'$ and $\rho_1(m) = \rho_2(m')$. Thus $m'$ is redundant, and we obtain the space on the right. Note that the composition $\rho_2 \circ i_{M'} = \rho_2 \circ \eta \circ s = i$, so we really do obtain $X_M$.  This then gives us a map from $X_M \to M \times_\mcB M'$ which we denote $\tau$. Explicitly, this is given by $\tau(x,m) = (m,i_{M'}(x))$.


Now, examine the push-pull combination $i^*_{M'} \pi_{2*}$ which appears above. It is equivalent to the composition $\rho_{1*} \tau^*$. Here we have abused notation slightly and used $\rho_1$ to denote the projection of $X_M = X \times_\mcB M$ to $B$. This map is the restriction of $\rho_1$ to $X_M$. The equivalence of these two operations can be seen because $i^*_{M'} \pi_{2*}$ integrates along $\theta$ and then restricts to the embedding of $X$ in $M'$ given by $i_{M'}$. On the other hand $\rho_{1*} \tau^*$ first restricts to $X_M$ and then integrates over $\theta$, giving the same answer.

We can now write the coupling as
$$
 \int_X \frac{i^*_{M'}\Ahat(M')}{\Ahat(N(X/M'))} e^{\beta}\, \rho_{1*}\tau^*\left(\frac{\Ahat\left(T(M/\mcB)\right)}{\sqrt{\Ahat(M)}} \pi_1^*(C) e^{\mcA' \wdg \mcA}\right)\ .
$$
We apply the projection formula along the fiber of $X_M$ giving
\begin{equation}
\label{tdbc}
 \int_{X_M} \rho_1^*\left(\frac{i^*_{M'}\Ahat(M')}{\Ahat(N(X/M'))} e^{\beta}\right) \tau^*\left(\frac{\Ahat\left(T(M/\mcB)\right)}{\sqrt{\Ahat(M)}}  \pi_1^*(C) e^{\mcA' \wdg \mcA}\right)\ .
\end{equation}
Notice that everything in the second set of parentheses is pulled back from $M$ except $e^{\mcA' \wdg \mcA}$. Since $\tau$ lifts $i_{M'}$ and $i_{M'}^*(\mcA') = \alpha$, we have
$$
\int_{X_M}  i_M^*\left(\frac{\Ahat\left(T(M/\mcB)\right)}{\sqrt{\Ahat(M)}}\right) \rho_1^*\left(\frac{i^*_{M'}\Ahat(M')}{\Ahat(N(X/M'))}e^{\beta}\right) C\  e^{\alpha \wdg \mcA}\ .
$$
On the level of cohomology, we have that
\begin{equation}
\label{ahatrels}
\frac{i_M^*\Ahat(M)}{\Ahat(N(X_M/M))}= i_M^*\Ahat\left(T(M/\mcB)\right) \rho_1^*\left(\frac{i^*_{M'}\Ahat(M')}{\Ahat(N(X/M'))}\right)\ .
\end{equation}
This can be easily seen, for example, from the following exact sequences:
\begin{eqnarray}
\label{embseq}
&0 \longrightarrow TX \longrightarrow i^*_{M'}TM' \longrightarrow N(X/M') \longrightarrow 0\ ,&\nn\\
&0 \longrightarrow TX_M \longrightarrow i^*_M TM \longrightarrow N(X_M/M) \longrightarrow 0\ ,&\\
&0 \longrightarrow i^*_MT(M/\mcB) \longrightarrow TX_M \longrightarrow \rho^*_1TX \longrightarrow 0\ .\nn&
\end{eqnarray}
If we assume that \eqref{ahatrels} holds on the level of forms as opposed to just cohomology classes, we have that the coupling of the T-dual RR-fields to the brane $X$ can be written as
$$
\int_{X_M} \frac{\sqrt{i^*_M\Ahat(M)}}{\Ahat(N(X_M/M))} e^{\beta} e^{\alpha \wdg \mcA} C\ ,
$$
which is exactly the coupling to the brane $X_M$ given by \eqref{xmcouple}. 

Of course, as we have emphasized, \eqref{ahatrels} will not be true on the level of forms for arbitrary connections on the various bundles. One can ask if there exists a set of connections on the five bundles above such that this relation holds. In fact, there does as we can see from the following bundle isomorphims:
\begin{eqnarray}
\label{bundisos}
&TM \cong \rho_1^*T\mcB \oplus T(M/\mcB)\ , \qquad TM' \cong \rho_2^*T\mcB \oplus T(M'/\mcB)\ ,& \nn\\
&N(X/M') \cong N(X/\mcB) \oplus i_{M'}^*T(M'/\mcB)\ ,&\\
&N(X_M/M) \cong \rho_1^*N(X/\mcB)\ .&\nn
\end{eqnarray}
The first two splittings are equivalent to the choice of a connection on the principal circle bundles $M$ and $M'$. The $T(M'/\mcB)$ subbundle of $N(X/M')$ represents the $U(1)$ action on $M'$ acting on the brane, and the splitting of $N(X/M')$ is induced from that on $TM'$. If we form the A-roof genera out of connections that respect these splittings, then the relation \eqref{ahatrels} holds. Nonetheless, as we will discuss in a moment, this is probably not how things are realized in physics.

While we are not claiming that the relations of the previous paragraph provide the physically correct answer, it is worth spending a moment to explore their consequences. In particular, these relations imply that the `correction' to the Buscher rules in \eqref{tdright} becomes
$$
\sqrt{\frac{\Ahat(M')}{\Ahat(M)}} \Ahat\left(T(M/\mcB)\right) = \sqrt{\Ahat\left(T(M/\mcB)\right)\Ahat\left(T(M'/\mcB)\right)}\ .
$$
A Riemannian structure and orientation on a bundle reduces the structure group to the special orthogonal group. If that respects the decompositions \eqref{bundisos}, the fact that $SO(1)$ is trivial means that the induced connection on a one dimensional component is also trivial. In particular, this implies that the correction we originally postulated in \eqref{tdright} is exactly equal to one. Thus, we are led to the conclusion that the Buscher rules are uncorrected and that every connection which appears in the brane coupling is pulled back from the base $\mcB$.

While these connections exist mathematically (although they may not be induced from the needed structures in the index theorems), they are unmotivated from the point of view of the physics. D-brane couplings should apply in all backgrounds whether or not there exists an Abelian symmetry, and, as such, the choice of the connection in the A-roof genus should be universal. Their exact form remains, then, a puzzle. There are a number of possible resolutions. The first is that the Buscher rules are, in fact, corrected. However, as we have seen, the simple correction \eqref{tdright} is not sufficient to ensure the invariance of the coupling; we need to further assume the relation \eqref{ahatrels}. Still, it is not too hard to add in correction terms by hand that render the full brane coupling invariant for any choice of forms representing the ratio of A-roof genera. However, any such correction (including that in \eqref{tdright}) almost inevitable spoils the self-duality derived above. This may not be a problem because of the subtleties associated with a self-dual field, but it is somewhat disturbing.

We can instead postulate that the Buscher rules are uncorrected (as we have seen follows from one attempt to satisfy the relation \eqref{ahatrels}). Note that since the low-energy effective action of type II string theories receives corrections only at $\mcO(\alpha'^3)$, it is natural to assume that the Buscher rules will not be corrected to up to that order.\footnote{That there are no corrections in the NSNS sector can further be seen by direct calculation \cite{Kaloper:1997ux}.} If we further assume that \eqref{tdright} is the correct expression, we have that
\begin{equation}
\label{nocorr}
\sqrt{\frac{\Ahat(M')}{\Ahat(M)}} \Ahat\left(T(M/\mcB)\right) = 1\ .
\end{equation}
A close examination will reveal that this is an extremely odd formula that seems difficult to satisfy unless all the relevant forms are pulled back from $\mcB$. Nonetheless, one might hope that there exist connections involving fields beyond the metric such that the equations of motion imply that \eqref{ahatrels} and \eqref{nocorr} hold. This would imply that T-duality only holds on-shell. One might go even further and require supersymmetry perhaps. Another possibility is that \eqref{tdright} is incorrect, and that the K-theoretic T-duality expression \eqref{tdkright} must be modified to keep the Buscher rules uncorrected. In another direction, it is possible that the coupling \eqref{xmcouple} is simply incomplete, and there exist further terms of a less topological character. One way to address these questions might be to understand the local form of the anomaly and demand its cancellation \cite{WIP}. One incomplete attempt in that direction is \cite{Scrucca:2000ae}. More generally, this can be resolved by an explicit calculation of the string scattering amplitudes \cite{WIP2} extending the work of \cite{Craps:1998tw,Craps:1998fn,Stefanski:1998yx}.

We do not have a solution to this puzzle and will devote the remainder of the paper to understanding how it relates to various properties one might expect T-duality to have in a K-theoretic context. Before proceeding, however, we note that one important consequence of \eqref{ahatrels} and \eqref{nocorr} is that
\begin{equation}
\label{aroofpull}
\frac{\sqrt{i_M^*\Ahat(M)}}{\Ahat(N(X_M/M))} = \rho_1^*\left(\frac{\sqrt{i^*_{M'}\Ahat(M')}}{\Ahat(N(X/M'))}\right)\ .
\end{equation}
As mentioned in the introduction, this means that the ratio of A-roof genera in the D-brane coupling is pulled back from $\mcB$.  This is important because it is not difficult to show that if it is not true, then T-duality implies that the coupling must contain additional terms which couple a D-brane to transverse RR-fields even for a single D-brane (in contrast to the Myers terms which are inherently non-Abelian). We will see in what follows that one can start with \eqref{aroofpull} and the assumption that the Buscher rules are uncorrected and derive the invariance of the D-brane coupling.

\section{The index pairing}\label{sec:index}

The goal of this section is to show that, when evaluated on differential K-theory classes invariant under the circle action, the index pairing is preserved by T-duality. In other words, given invariant elements $H \in \check{H}^3(M)$, $\mcC \in \check{K}^\bullet_{-H}(M)$, $\mcD \in \check{K}^\bullet_{H}(M)$ and their primed T-duals, we want that
\begin{equation}
\label{ipinv}
\int_M \mcC \cdot \mcD = \int_{M'} \mcC' \cdot \mcD'\ .
\end{equation}
For this calculation, we will assume that the Buscher rules are uncorrected \eqref{nocorr}, but \textit{not} that relation \eqref{ahatrels} holds. Following \eqref{fmchern}, let us define
\begin{equation}
\label{cchern}
\mathrm{CH}(\mcC) \defeq \sqrt{\Ahat(M)} \mathrm{ch}(\mcC)\ ,
\end{equation}
and similarly for any differential K-theory class. This expression is chosen so that the pairing in cohomology is the same as the pairing in K-theory. In other words, we have
$$
\int_M \mathrm{CH}(\mcC)\wdg \mathrm{CH}(\mcD) = \int_M \Ahat(M) \mathrm{ch}(\mcC) \mathrm{ch}(\mcD) = \int_M \mcC \cdot \mcD\ .
$$

Now, given a form $C$, let us denote the Buscher T-dual by
$$
T(C) \defeq \pi_{2*}\left(e^{\mcA' \wdg\mcA} \pi_1^*(D)\right)\ .
$$
The assumption that the Buscher rules are uncorrected is equivalent to the statement that
\begin{equation}
\label{tchcomm}
T(\mathrm{CH}(C)) = \mathrm{CH}(\mcC')\ ,
\end{equation}
where $\mcC'$ is some K-theoretic version of T-duality along the lines of \eqref{tdkright}. 

We will show that
\begin{equation}
\label{fpinv}
\int_M C \wdg D = \int_{M'} T(C) \wdg T(D)\ .
\end{equation}
Then \eqref{ipinv} follows from \eqref{tchcomm} and  \eqref{fpinv} by
$$
\int_M \mcC \cdot \mcD = \int_M \mathrm{CH}(\mcC)\wdg \mathrm{CH}(\mcD) = \int_{M'} T(\mathrm{CH}(\mcC))\wdg T(\mathrm{CH}(\mcD)) = \int_{M'} \mcC' \cdot \mcD'\ .
$$

Since the forms in \eqref{fpinv} are invariant under the circle action, we can define
$$
C = c_1 + c_2 \wdg \mcA\ , \qquad D = d_1 + d_2 \wdg \mcA\ ,
$$
where $c_1,c_2,d_1$ and $d_2$ are all implicitly pulled back from $\mcB$. Then, for the right hand side of \eqref{fpinv}, we have:
\begin{eqnarray*}
\int_{M'}T(C) T(D) &=& \int_{M'}  \pi_{2*}\left(e^{\mcA \wdg \mcA'}\pi_1^*C\right) \pi_{2*}\left(e^{\mcA' \wdg \mcA} \pi_1^*C\right)\\
&=&  \int_{M \times_\mcB M'}e^{\mcA \wdg \mcA'}\pi_1^*C\ \pi_2^*\pi_{2*}\left(e^{\mcA' \wdg \mcA} \pi_1^*D\right) \\
&=& \int_{M \times_\mcB M'}e^{\mcA \wdg \mcA'}\pi_1^*C\ \pi_2^*\left(d_2 + d_1 \wdg \mcA'\right) \\
&=& \int_{\mcB}  (c_1 d_2 + (-1)^{d_1} c_2 d_1)\ .
\end{eqnarray*}
On the other hand, the left hand side of \eqref{fpinv} is given by
$$
\int_M  (c_1 + c_2 \wdg \mcA) (d_1 + d_2 \wdg \mcA) = \int_\mcB  ( c_1 d_2 + (-1)^{d_1}c_2 d_1)\ .
$$
Thus, we see that they are equal, and we are done. 

\section{T-duality of branes}\label{sec:branes}

In section \ref{sec:setup}, we treated the T-dual D-brane in an \textit{ad hoc} manner. Given that RR-fields measure the charges of D-branes, however, one expects that the T-duality of branes should follow from a geometric formula such as \eqref{tdwrong} or its K-theoretic generalization \eqref{tdkright}. That is the goal of this section.

We saw in section \ref{sec:ktheory} that the differential K-theory class of a D-brane is defined by the K-theoretic pushforward $i_{M*}(1)$. After taking (modified) Chern characters, we obtain the differential form
$$
\eta_{X_M}\defeq \sqrt{\Ahat(M)} i_{M*}\!\left(\Ahat(N(X_M/M))^{-1}e^{B}\right)\ .
$$
Recall that the D-branes are twisted by $-H$. We will assume as in \eqref{tchcomm} that T-duality and the modified Chern character commute. Thus, we have
$$
\eta'_{X_M} \defeq T(\eta_{X_M}) = \pi_{2*}\!\left(e^{\mcA \wdg \mcA'} \pi_1^* i_{M*}\!\left(\frac{\sqrt{i_M^*\Ahat(M)}}{\Ahat(N(X_M/M))}e^{B}\right)\right)
$$
On the other hand, we have the T-dual brane as defined in section \ref{sec:setup}
$$
\eta_{X} \defeq  \sqrt{\Ahat(M')}i_{M'*}\!\left(\Ahat(N(X/M'))^{-1}e^{B'}\right)\ .
$$
We are now faced with the question: are $\eta'_{X_M}$ and $\eta_{X}$ equal? Unfortunately, the answer is no. How, then, do we reconcile this with T-duality? To motivate the answer, recall that these are not differential forms but are instead currents. They are the analogue of delta functions in the world of differential forms and are defined similarly to distributions as elements in the dual vector space to some nice space of differential forms.\footnote{Even more, we should really view these as Chern characters of elements in currential K-theory. Since we are choosing the modified Chern character, the pairing in K-theory is the same as that in cohomology, however, and we can ignore this point.} From the point of view of the physics, what this means is that all the questions we want to answer involve the integrals of the form
$$
\int_{M'}  i_{M'*}(\sigma)\  \omega\ ,
$$
where $\sigma$ is a form on $X$ and $\omega$ a form on $M'$. The definition of the current $i_{M'*}(\sigma)$ is that this expression is equivalent to
$$
\int_X \sigma\ i_{M'}^*\!\!\left(\omega\right)\ ,
$$
which we can easily evaluate. If two currents agree on all possible `test forms' $\omega$, then they are equal.

As stated above, it is easily verified that this is false for $\eta'_{X_M}$ and $\eta_X$, so the currents are unequal. We can now understand the problem, however. T-duality only holds if the RR-fields are invariant with respect to the circle action. In other words, we will only be testing our currents against differential forms invariant under that action. Let us choose such a form $\omega$. Because of its invariance we can write it as $\omega = \omega_1 + \omega_2 \wdg \mcA'$. To begin with, for $\eta_X$ we have
\begin{equation}
\label{eta1}
\int_{M'} \eta_X \wdg \omega = \int_X \frac{\sqrt{i_{M'}^*\Ahat(M')}}{\Ahat(N(X/M')}e^{B'} i_{M'}^* \omega =  \int_X \frac{\sqrt{i_{M'}^*\Ahat(M')}}{\Ahat(N(X/M')} e^\beta \left( i^*\omega_1 + i^*\omega_2 \wdg \alpha\right)\ .
\end{equation}
Here, we have used $B' = \beta$ from section \ref{sec:setup}.

For $\eta'_{X_M}$, we evaluate
\begin{equation}
\label{eta2}
\begin{split}
\int_{M'} \eta'_{X_M} \wdg \omega &= \int_{M'}  \pi_{2*}\!\left(e^{\mcA \wdg \mcA'} \pi_1^* i_{M*}\!\left(\frac{\sqrt{i_M^*\Ahat(M)}}{\Ahat(N(X_M/M))}e^{B}\right)\right) \omega\\
&= \int_{M \times_\mcB M'} e^{\mcA \wdg \mcA'} \pi_1^* i_{M*}\!\left(\frac{\sqrt{i_M^*\Ahat(M)}}{\Ahat(N(X_M/M))}e^{B}\right)\pi_2^*\left(\omega\right) \\
&= \int_M   i_{M*}\left(\frac{\sqrt{i_M^*\Ahat(M)}}{\Ahat(N(X_M/M))}e^{B}\right) \pi_{1*}\left(e^{\mcA \wdg \mcA'} \pi_2^*\left(\omega\right)\right)\\
&= \int_{X_M}\frac{\sqrt{i_M^*\Ahat(M)}}{\Ahat(N(X_M/M))}e^{B} i_M^*\pi_{1*}\left(e^{\mcA \wdg \mcA'} \pi_2^*\left(\omega\right)\right)\ .
\end{split}
\end{equation}

We want to integrate over the fiber to obtain something on $X$. Ignoring, for the moment, the terms involving the A-roof genus, we have
$$
\rho_{1*}\left(e^{B} i_M^*\pi_{1*}\left(e^{\mcA \wdg \mcA'} \pi_2^*(\omega)\right) \right) =
\rho_{1*}\left(e^{B} i_M^*\left(\omega_2 + \omega_1 \wdg \mcA\right)\right) = e^\beta\left( i^*\omega_1+ i^* \omega_2 \wdg\alpha\right)\ ,
$$
making use of the definition $B = \beta - \mcA \wdg \alpha$ from section \ref{sec:setup}.

Comparing \eqref{eta1} and \eqref{eta2}, we see that they are equal when
$$
\frac{\sqrt{i_M^*\Ahat(M)}}{\Ahat(N(X_M/M))} = \rho_1^*\left(\frac{\sqrt{i^*_{M'}\Ahat(M')}}{\Ahat(N(X/M'))}\right)\ .
$$
This is precisely equation \eqref{aroofpull} above.

In light of this result and that of the previous section, we can now rederive the invariance of the coupling shown in section \ref{sec:verif}. Recall that the coupling can be expressed as the K-theoretic integral \eqref{ktint}
$$
\int_M i_{M*}(1) \cdot \mcC\ .
$$
As we have established that the T-dual of $i_{M*}(1)$ is $i_{M'*}(1)$ when evaluated on invariant classes, \eqref{ipinv} implies that
$$
\int_M i_{M*}(1) \cdot \mcC = \int_{M'} (i_{M*}(1))' \cdot \mcC' = \int_{M'} i_{M'*}(1) \cdot \mcC'\ ,
$$
which establishes that the coupling is invariant under T-duality. However, this proof gives a somewhat different perspective than that in section \ref{sec:verif} because rather than assuming \eqref{ahatrels}, we have instead assumed that the Buscher rules are uncorrected \eqref{tchcomm} and that the A-roof corrections in the coupling are pulled back from $\mcB$ \eqref{aroofpull}. Understanding the correct perspective will have to be the subject of future work.

\acknowledgments

We thank Mohammed Garousi and Sav Sethi for many helpful discussions. AB also thanks Ben Craps, Jacques Distler, Dan Freed, Jeff Harvey, Alexander Kahle, Ruben Minasian, Greg Moore, Rob Myers, and Alessandro Tomasiello for discussions and e-mails. We thank the the Kavli Institute for Theoretical Physics for their hospitality and support while most of this work was being accomplished. KB also thanks the Aspen Center for Physics for their hospitality. This work is supported by the NSF under grants PHY05-05757, PHY05-51164 and PHY05-55575 and by Texas A\&M University.

\bibliographystyle{jhep}
\bibliography{thebib}

\providecommand{\href}[2]{#2}\begingroup\raggedright\begin{thebibliography}{10}

\bibitem{Giveon:1994fu}
A.~Giveon, M.~Porrati, and E.~Rabinovici, {\it {Target space duality in string
  theory}},  {\em Phys. Rept.} {\bf 244} (1994) 77--202,
  [\href{http://xxx.lanl.gov/abs/hep-th/9401139}{{\tt hep-th/9401139}}].

\bibitem{Buscher:1987qj}
T.~H. Buscher, {\it {Path Integral Derivation of Quantum Duality in Nonlinear
  Sigma Models}},  {\em Phys. Lett.} {\bf B201} (1988) 466.

\bibitem{Buscher:1987sk}
T.~H. Buscher, {\it {A Symmetry of the String Background Field Equations}},
  {\em Phys. Lett.} {\bf B194} (1987) 59.

\bibitem{Buscher:1985kb}
T.~H. Buscher, {\it {Quantum Corrections and Extended Supersymmetry in New
  Sigma Models}},  {\em Phys. Lett.} {\bf B159} (1985) 127.

\bibitem{Meessen:1998qm}
P.~Meessen and T.~Ortin, {\it {An Sl(2,Z) multiplet of nine-dimensional type II
  supergravity theories}},  {\em Nucl. Phys.} {\bf B541} (1999) 195--245,
  [\href{http://xxx.lanl.gov/abs/hep-th/9806120}{{\tt hep-th/9806120}}].

\bibitem{Bergshoeff:1995as}
E.~Bergshoeff, C.~M. Hull, and T.~Ortin, {\it {Duality in the type II
  superstring effective action}},  {\em Nucl. Phys.} {\bf B451} (1995)
  547--578, [\href{http://xxx.lanl.gov/abs/hep-th/9504081}{{\tt
  hep-th/9504081}}].

\bibitem{Hassan:1999bv}
S.~F. Hassan, {\it {T-duality, space-time spinors and R-R fields in curved
  backgrounds}},  {\em Nucl. Phys.} {\bf B568} (2000) 145--161,
  [\href{http://xxx.lanl.gov/abs/hep-th/9907152}{{\tt hep-th/9907152}}].

\bibitem{Hassan:1999mm}
S.~F. Hassan, {\it {SO(d,d) transformations of Ramond-Ramond fields and space-
  time spinors}},  {\em Nucl. Phys.} {\bf B583} (2000) 431--453,
  [\href{http://xxx.lanl.gov/abs/hep-th/9912236}{{\tt hep-th/9912236}}].

\bibitem{Hori:1999me}
K.~Hori, {\it {D-branes, T-duality, and index theory}},  {\em Adv. Theor. Math.
  Phys.} {\bf 3} (1999) 281--342,
  [\href{http://xxx.lanl.gov/abs/hep-th/9902102}{{\tt hep-th/9902102}}].

\bibitem{Myers:1999ps}
R.~C. Myers, {\it {Dielectric-branes}},  {\em JHEP} {\bf 12} (1999) 022,
  [\href{http://xxx.lanl.gov/abs/hep-th/9910053}{{\tt hep-th/9910053}}].

\bibitem{Green:1996dd}
M.~B. Green, J.~A. Harvey, and G.~W. Moore, {\it {I-brane inflow and anomalous
  couplings on D-branes}},  {\em Class. Quant. Grav.} {\bf 14} (1997) 47--52,
  [\href{http://xxx.lanl.gov/abs/hep-th/9605033}{{\tt hep-th/9605033}}].

\bibitem{Cheung:1997az}
Y.-K.~E. Cheung and Z.~Yin, {\it {Anomalies, branes, and currents}},  {\em
  Nucl. Phys.} {\bf B517} (1998) 69--91,
  [\href{http://xxx.lanl.gov/abs/hep-th/9710206}{{\tt hep-th/9710206}}].

\bibitem{Witten:1998cd}
E.~Witten, {\it D-branes and {K}-theory},  {\em JHEP} {\bf 12} (1998) 019,
  [\href{http://xxx.lanl.gov/abs/hep-th/9810188}{{\tt hep-th/9810188}}].

\bibitem{Moore:1999gb}
G.~W. Moore and E.~Witten, {\it Self-duality, ramond-ramond fields, and
  k-theory},  {\em JHEP} {\bf 05} (2000) 032,
  [\href{http://xxx.lanl.gov/abs/hep-th/9912279}{{\tt hep-th/9912279}}].

\bibitem{Bouwknegt:2003vb}
P.~Bouwknegt, J.~Evslin, and V.~Mathai, {\it {T-duality: Topology change from
  H-flux}},  {\em Commun. Math. Phys.} {\bf 249} (2004) 383--415,
  [\href{http://xxx.lanl.gov/abs/hep-th/0306062}{{\tt hep-th/0306062}}].

\bibitem{Freed:2000ta}
D.~S. Freed, {\it Dirac charge quantization and generalized differential
  cohomology},  in {\em Surveys in differential geometry}, Surv. Differ. Geom.,
  VII, pp.~129--194.
\newblock Int. Press, Somerville, MA, 2000.
\newblock \href{http://xxx.lanl.gov/abs/hep-th/0011220}{{\tt hep-th/0011220}}.

\bibitem{Becker:2009zx}
K.~Becker, C.~Bertinato, Y.-C. Chung, and G.~Guo, {\it {Supersymmetry breaking,
  heterotic strings and fluxes}},
  \href{http://xxx.lanl.gov/abs/0904.2932}{{\tt 0904.2932}}.

\bibitem{dfpcomm}
D.~S. Freed \textit{private communication}.

\bibitem{Belov:2006xj}
D.~M. Belov and G.~W. Moore, {\it {Type II actions from 11-dimensional
  Chern-Simons theories}},  \href{http://xxx.lanl.gov/abs/hep-th/0611020}{{\tt
  hep-th/0611020}}.

\bibitem{Minasian:1997mm}
R.~Minasian and G.~W. Moore, {\it {K-theory and Ramond-Ramond charge}},  {\em
  JHEP} {\bf 11} (1997) 002,
  [\href{http://xxx.lanl.gov/abs/hep-th/9710230}{{\tt hep-th/9710230}}].

\bibitem{Freed:2009it}
D.~S. Freed and J.~Lott, {\it An index theorem in differential k-theory},
  \href{http://xxx.lanl.gov/abs/0907.3508}{{\tt 0907.3508}}.

\bibitem{Carey:2007dt}
A.~L. Carey, J.~Mickelsson, and B.-L. Wang, {\it Differential twisted k-theory
  and applications},  \href{http://xxx.lanl.gov/abs/0708.3114}{{\tt
  0708.3114}}.

\bibitem{Craps:1998tw}
B.~Craps and F.~Roose, {\it {(Non-)anomalous D-brane and O-plane couplings: The
  normal bundle}},  {\em Phys. Lett.} {\bf B450} (1999) 358,
  [\href{http://xxx.lanl.gov/abs/hep-th/9812149}{{\tt hep-th/9812149}}].

\bibitem{Carey:2005mx}
A.~L. Carey and B.-L. Wang, {\it Thom isomorphism and push-forward map in
  twisted {$K$}-theory},  {\em J. K-Theory} {\bf 1} (2008), no.~2 357--393,
  [\href{http://xxx.lanl.gov/abs/math.KT/0507414}{{\tt math.KT/0507414}}].

\bibitem{Kaloper:1997ux}
N.~Kaloper and K.~A. Meissner, {\it {Duality beyond the first loop}},  {\em
  Phys. Rev.} {\bf D56} (1997) 7940--7953,
  [\href{http://xxx.lanl.gov/abs/hep-th/9705193}{{\tt hep-th/9705193}}].

\bibitem{WIP}
 Work in progress.

\bibitem{Scrucca:2000ae}
C.~A. Scrucca and M.~Serone, {\it {A note on the torsion dependence of D-brane
  RR couplings}},  {\em Phys. Lett.} {\bf B504} (2001) 47--54,
  [\href{http://xxx.lanl.gov/abs/hep-th/0010022}{{\tt hep-th/0010022}}].

\bibitem{WIP2}
 Work in progress.

\bibitem{Craps:1998fn}
B.~Craps and F.~Roose, {\it {Anomalous D-brane and orientifold couplings from
  the boundary state}},  {\em Phys. Lett.} {\bf B445} (1998) 150--159,
  [\href{http://xxx.lanl.gov/abs/hep-th/9808074}{{\tt hep-th/9808074}}].

\bibitem{Stefanski:1998yx}
B.~Stefanski{,}~Jr., {\it {Gravitational couplings of D-branes and O-planes}},
  {\em Nucl. Phys.} {\bf B548} (1999) 275--290,
  [\href{http://xxx.lanl.gov/abs/hep-th/9812088}{{\tt hep-th/9812088}}].

\end{thebibliography}\endgroup

\end{document}